\documentclass[letter]{aa}  
\usepackage{graphicx}
\usepackage[T1]{fontenc}
\usepackage{txfonts}
\usepackage{natbib}

\bibpunct{(}{)}{;}{a}{}{,}
\begin{document}
   \title{The limb-darkened Arcturus}
   \subtitle{Imaging with the IOTA/IONIC interferometer}

   \author{S.~Lacour
          \inst{1,2,3}
	  \and
          S.~Meimon\inst{4}
          \and
          E.~Thi\'ebaut\inst{5}
          \and
          G.~Perrin\inst{1}
          \and
          T.~Verhoelst\inst{6,7}\fnmsep \thanks{Postdoctoral Fellow of the Fund
for Scientific Research,
                Flanders}
          \and
          E.~Pedretti\inst{8} 
          \and 
          P.~A.~Schuller\inst{9,10} 
          \and 
          L.~Mugnier\inst{4}
	  \and
          J.~Monnier\inst{11}
          \and
          J.P.~Berger\inst{3}
	  \and
          X.~Haubois\inst{1}
          \and
          A.~Poncelet\inst{1}
	  \and
          G.~Le~Besnerais\inst{4}
          \and
          K.~Eriksson\inst{12}
          \and
          R. Millan-Gabet\inst{13}
	  \and
	  M.~Lacasse\inst{9}
	  \and 
	  W.~Traub\inst{9,14}
}

   \offprints{S. Lacour}

   \institute{ Observatoire de Paris, LESIA, CNRS/UMR\,8109, 92190 Meudon,
     France 
     \and Sydney University, School of Physics, N.S.W. 2006,
     Australia 
     \and Laboratoire d'Astrophysique de Grenoble, CNRS/UMR\,5571,
     38041 Grenoble, France 
     \and Office National d'\'Etudes et de
     Recherches A\'eronautiques, DOTA, 92322 Chatillon, France 
     \and
     Centre de Recherche Astrophysique de Lyon, CNRS/UMR\,5574, 
     69561 Saint Genis Laval, France 
     \and Instituut voor Sterrenkunde, K.U. Leuven, 3001
     Leuven, Belgium 
     \and University of Manchester, Jodrell Bank Centre for Astrophysics,
     Manchester, M13 9PL, U.K.
     \and University of St Andrews, North Haugh,
     St Andrews,  KY16 9SS, Scotland, UK
     \and Harvard-Smithsonian Center for
     Astrophysics, Cambridge, MA, USA 
     \and Institut d'Astrophysique
     Spatial, Universit\'e Paris-Sud, 91405 Orsay, France 
     \and Department of Astronomy, University of
     Michigan, Ann Arbor, MI, USA 
     \and
     Department of Astronomy and Space Physics, Uppsala University,
     75120 Uppsala, Sweden 
     \and Caltech/Michelson Science Center,
     Pasadena, CA, USA
     \and
     Jet Propulsion Lab,  M/S 301-451, 4800 Oak Grove Dr., Pasadena CA, 91109
}


  \abstract 
{} 
{ This paper is an H band interferometric examination of
  Arcturus, a star frequently used as a spatial and spectral
  calibrator.}
{ Using the IOTA 3 telescope interferometer, we performed
  spectro-interferometric observation ($R\approx35$) of
  Arcturus. Atmospheric models and prescriptions were fitted to the
  data to derive the brightness distribution of the photosphere. Image
  reconstruction was also obtained using two software algorithms:
  \textsc{Wisard} and \textsc{Mira}.}
{ An achromatic power
  law proved to be a good model of the brightness distribution, with a
  limb darkening compatible with the one derived from atmospheric
  model simulations using our \textsc{marcs} model.  A Rosseland diameter of $21.05\pm0.21$ was derived,
  corresponding to an effective temperature of $T_{\rm
    eff}=4295\pm26$\,K. No companion was detected from the closure
  phases, with an upper limit on the brightness ratio of
  $8\times10^{-4}$ at 1AU. Dynamic range at such distance from the
  photosphere was established at $1.5\times10^{-4}$
  ($1\sigma\,$rms). An upper limit of $1.7\times10^{-3}$ was also
  derived for the level of brightness asymmetries present on the
  photosphere.  }
{}

   \maketitle
%

\section{Introduction}

Arcturus' curse is to be too popular. Many new generations of
instruments -- including interferometers -- observe it as a test
object. The reasons for that: this red giant is bright, large, and
spectrally well-defined. The curse is: since the instruments are new,
the star usually takes face on the bizarre systematic errors of
challenging observations. We can cite, among others, inexact diameter
and temperature measurements (prompting \citet{1999AJ....117.2998G} to
write an article entitled ``The Effective Temperature of Arcturus''),
or false duplicity observations \citep[``Arcturus as a Double Star''
  by][]{1998Obs...118..299G}.

This paper does not pretend to be an exception -- interferometry is
still a challenging technique. The main difference is in the
interferometer used: at the time of our observations, IOTA was more
wise than new \citep[we shall note that Arcturus has already been
  observed several times by IOTA and led to three different
  publications;][]{1996AJ....111.1705D,1998A&A...331..619P,2005A&A...435..289V}.
The initial goal of a new observation run was to leverage the more
extended capability of IOTA to show the ability of the interferometer
to do a reliable image of a commonly observed object.

Indeed, image reconstruction is difficult. Even though it is routinely
performed by the current generation of radio interferometers, this
technique --  herein called ``regularized imaging'' -- remains marginal in optical interferometry. This is simply
due to a lack of spatial frequency coverage. Optical interferometers are
usually more difficult to build, and the complexity quickly increases
with the number of telescopes. Therefore, since the amount of
information accessible in the Fourier plane is sparse, our ability to
reconstruct a reliable image of a complex object is limited. 

A more common data analysis is to suppose the object to be conform to
a model -- or prescription. Originally, it consisted in fitting
visibility curves of uniform disks \citep[e.g.][]{1921ApJ....53..249M,
  1986A&A...166..204D}. With time, it included more complicated
models, e.g. limb-darkened disks
\citep[e.g.][MkIII]{1996A&A...312..160Q}, disks with spots
\citep[e.g.][COAST]{2000MNRAS.315..635Y}, disks with a molecular
envelope \citep[e.g.][IOTA]{2004A&A...426..279P}, etc.  Both
  methods, regularized imaging and model fitting, complement each
  other. The role of regularized imaging is commonly to guide the
  choice and complexity of a model. The role of model fitting is to
  obtain the highest precision results on the parameters of the
  model. The pitfall may be when the model does not best suit the
  object, hence the need for quality regularization imaging.

Recently the maturation of interferometric facilities (e.g. IOTA,
CHARA, VLTI) reached the point where $u$-$v$ coverage (both in
amplitude and phase) allows regularized imaging
\citep{2007arXiv0706.0867M,lacour_these}.  Here we present the data on
Arcturus, used as a test star for optical interferometry
reconstruction softwares. Several astrophysical issues also justify
this investigation: 1) what is the limb darkening? Is it compatible
with red giant atmosphere modeling
\citep{2000MNRAS.318..387D,2000A&A...363.1081C}? 2) are the disputed
previous detections of a companion compatible with our observations
\citep{1997A&A...323L..49P,2005A&A...435..289V,2007PASP..119..237B}?

The outline of this paper is as follows. Section~\ref{sc:obs} gives an
overview of the IOTA interferometer, describes the data reduction
process and shortly present the dataset.  Section~\ref{sc:Com_model}
compares our data with atmosphere models, using either limb-darkening
prescriptions or a more evolved atmospheric simulation (the
\textsc{Marcs} model). Section~\ref{sc:comp} investigates on a
possible deviation from point symmetry. Results of image
reconstruction are presented in Section~\ref{sc:image}, and
Section~\ref{sc:conclusion} conclude.


\section{Observation and data reduction} \label{sc:obs}

\subsection{Description of IOTA observations}

   \begin{figure}
   \centering
   \includegraphics[scale=.6]{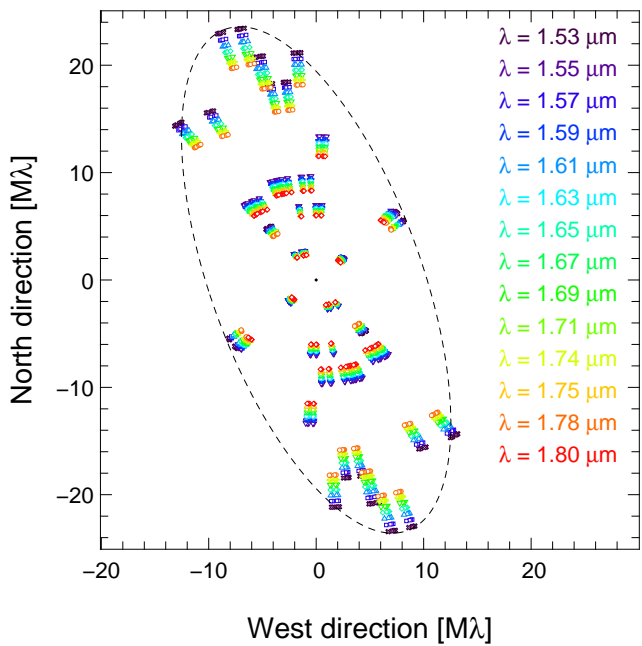}
      \caption{$u$-$v$ coverage. Maximum projected baseline length
              is 37.7 meters. The lack of high frequency information
              in the East-West direction is due to the geometry of
              IOTA.}
         \label{fig:UV_planes}
   \end{figure}

   \begin{figure*}
   \centering
   \includegraphics[width=17cm]{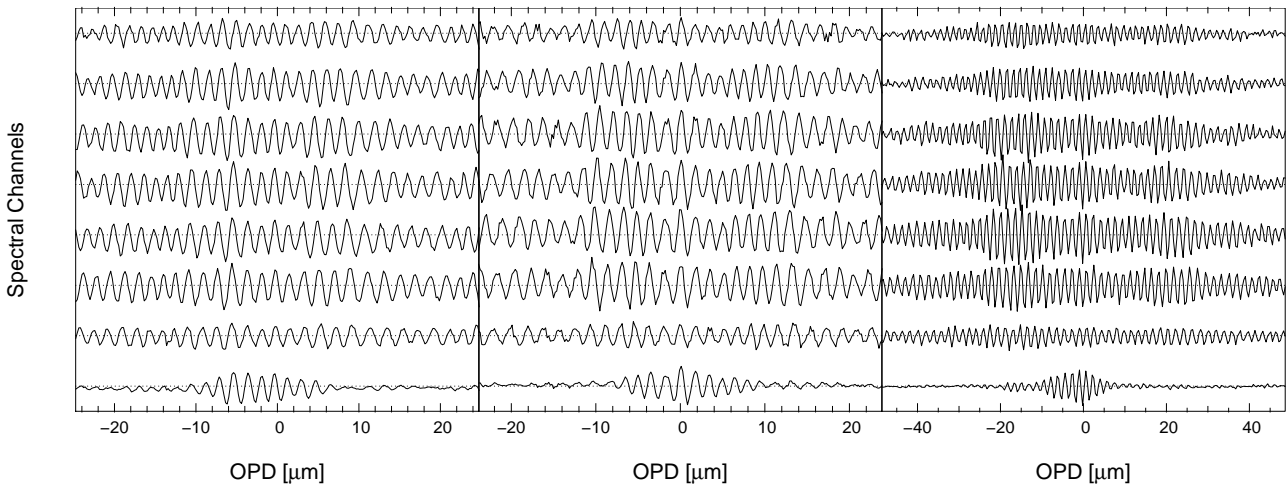}
      \caption{Single scan obtained on calibrator HD\,120477. It shows
        the flux (arbitrary units) as a function of the optical path
        difference (OPD) for the three baselines, and the seven
        spectral channels. Eye inspection allows to observe the
        decreasing frequency -- and therefore increasing wavelength --
        from top to bottom. The bottom fringes correspond to the sum
        of the spectral channels, showing a radical change in the
        coherence length. Each datapoint is composed of 200 scans.}
         \label{fig:fringes}
   \end{figure*}

The interferometric data presented herein were obtained using the IOTA
(Infrared-Optical Telescope Array) interferometer
\citep{2003SPIE.4838...45T}, a long baseline interferometer which
operates at near-infrared wavelengths. It consists of three 0.45 meter
telescopes movable among 17 stations along two orthogonal linear
arms. IOTA synthesizes a total aperture size of $35 \times 15\,$m,
corresponding to an angular resolution of $\approx 10 \times 23$
milliarcseconds at 1.65 $\mu$m.  Visibility and closure phase
measurements were obtained using the integrated optics combiner IONIC
\citep{2003SPIE.4838.1099B}; light from the three telescopes is
focused into single-mode fibers and injected into the planar
integrated optics (IO) device. Six IO couplers allow recombinations
between each pair of telescopes. Fringe detection is done using a
Rockwell PICNIC detector \citep{2004PASP..116..377P}. The interference
fringes are temporally-modulated on the detector by scanning piezo
mirrors placed in two of the three beams of the interferometer.

Observations were undertaken in the H band
($1.5\,\mu$m$\,\leq\,\lambda\leq\,1.8\,\mu$m) divided into 7 spectral
channels.  The science target observations are interleaved with
identical observations of unresolved or partially resolved stars, used
to calibrate the interferometer's instrumental response and effects of
atmospheric seeing on the visibility amplitudes. The calibrator
sources were chosen in two different catalogs:
\citet{2002A&A...393..183B} and \citet{2006A&A...447..783M}; using
criteria on the separation ($\lesssim 10$ degrees) and magnitude. The
calibrators are listed in Table~\ref{tb:calib}.

\begin{table}
\caption{Calibrators}
\label{tb:calib}
\centering
\begin{tabular}{lccc}
\hline
\hline
&Calibrator & Spectral Type & UD diameter \\
\hline
1&HD\,120477 &    K5.5\,III   & $4.460 \pm  0.050$ \\
2&HD\,125560 &    K3\,III     & $1.910  \pm 0.021$ \\
3&HD\,129972&    G8.5\,III   & $1.540  \pm 0.020$ \\
\hline
\end{tabular}
\end{table}

Arcturus was observed in May 2006 during 5 nights and using 5 different
configurations of the interferometer. Full observation information can
be found in Table~\ref{tb:log}, including dates of observation,
interferometer configurations and calibrators.
Fig.~\ref{fig:UV_planes} shows the $u$-$v$ coverage achieved during this
observation run. The geometry of the IOTA interferometer and the
position of the star on the sky constrained the extent of frequency
coverage. We covered a frequency range equivalent to the one of an
elliptical telescope of aperture $38\times15$ meters, with a 20
degree inclination East of North.

\begin{table}
\caption{Arcturus observing log}
\label{tb:log}
\centering
\begin{tabular}{ccl}
\hline \hline 
Date & Interferometer & \multicolumn{1}{c}{Calibrator} \\ 
(UT) & Configuration$^{\mathrm{a}}$  & \multicolumn{1}{c}{(Table~\ref{tb:calib})} \\ 
\hline 
2006 May 11 & A15-B05-C10 & 1, 2, 3\\
2006 May 12 & A15-B05-C00 & 1, 2, 3\\
2006 May 13 & A15-B15-C00 & 1, 3\\
2006 May 14 & A30-B15-C00 & 2\\
2006 May 16 & A35-B15-C25 & 2, 3\\
\hline
\end{tabular}
\begin{list}{}{}
\item[$^{\mathrm{a}}$] Configuration refers to the location in meters
  of telescopes A, B, C on the NE, SE and NE arms respectively.
\end{list}
\end{table}

\subsection{Data reduction} \label{sc:data_red}

Reduction of the IONIC visibility data was carried out using custom
software similar in its main principles to the one described by
\citet{1997A&AS..121..379C}. We measured the power spectrum of each
interferogram (proportional to the target squared visibility, $V^2$),
after correcting for intensity fluctuations and subtracting bias terms
from read noise, residual intensity fluctuations, and photon noise
\citep{2003A&A...398..385P}. Next, the data pipeline applies a
correction for the variable flux ratios for each baseline by using a
flux transfer matrix \citep{2001PASP..113..639M}. Finally, raw squared
visibilities are calibrated using the raw visibilities obtained by the
same means on the calibrator stars. Calibration accuracy had been
studied by extensive observation of the binary star $\lambda$ Vir. For
bright stars (H mag $\lesssim 5$), \citet{2007ApJ...659..626Z} have
validated 2 \% calibration error for $V^2$; corresponding to a 1\%
error in visibility. We therefore systematically added a 2\%
calibration error on all the squared visibilities present in this
paper.

In order to measure the closure phase (CP), a fringe tracking
algorithm was applied in real-time while recording interferograms
\citep{2005ApOpt..44.5173P}, ensuring that interference occurs
simultaneously for all baselines. We required that interferograms are
detected for at least two of the three baselines in order to assure a
good closure phase measurement. This technique, called
``baseline bootstrapping'' allowed precise visibility and closure
phase measurements for a third baseline with very small coherence
fringes.  We followed the method of \citet{1996A&A...306L..13B} for
calculating the complex triple amplitude and deriving the closure
phase. Pair-wise combiners (such as IONIC) can have a large
instrumental offset for the closure phase which requires to be
calibrated by the closure phase of the calibrator stars. We noticed
very stable closure phase measurements during the nights with drifts
of less than a degree. 

\subsection{Wavelength calibration}
\label{sc:wave}

   \begin{figure}
   \centering
   \includegraphics[width=8.cm]{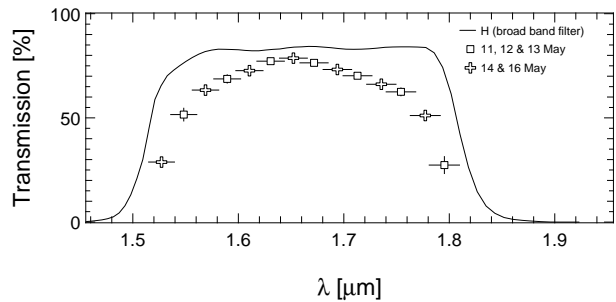}
      \caption{Relative photometry between the different spectral
      channels as a function of the wavelength (arbitrary vertical
      units). The wavelength was determined by measuring the frequency
      of the fringes as shown in Fig.~\ref{fig:fringes}.}
         \label{fig:wav}
   \end{figure}

   \begin{figure}
   \centering
   \includegraphics[scale=.47]{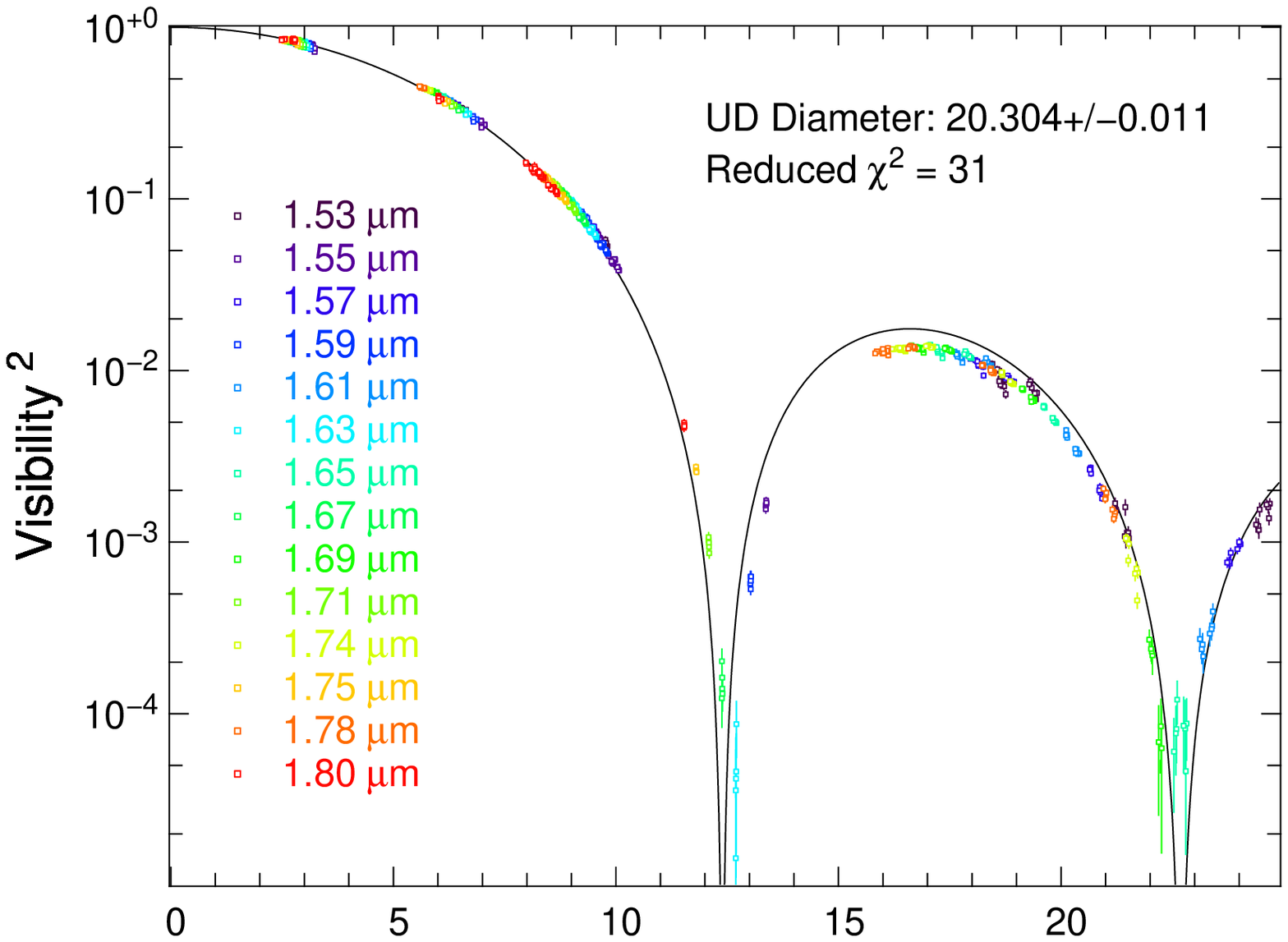}
   \includegraphics[scale=.47]{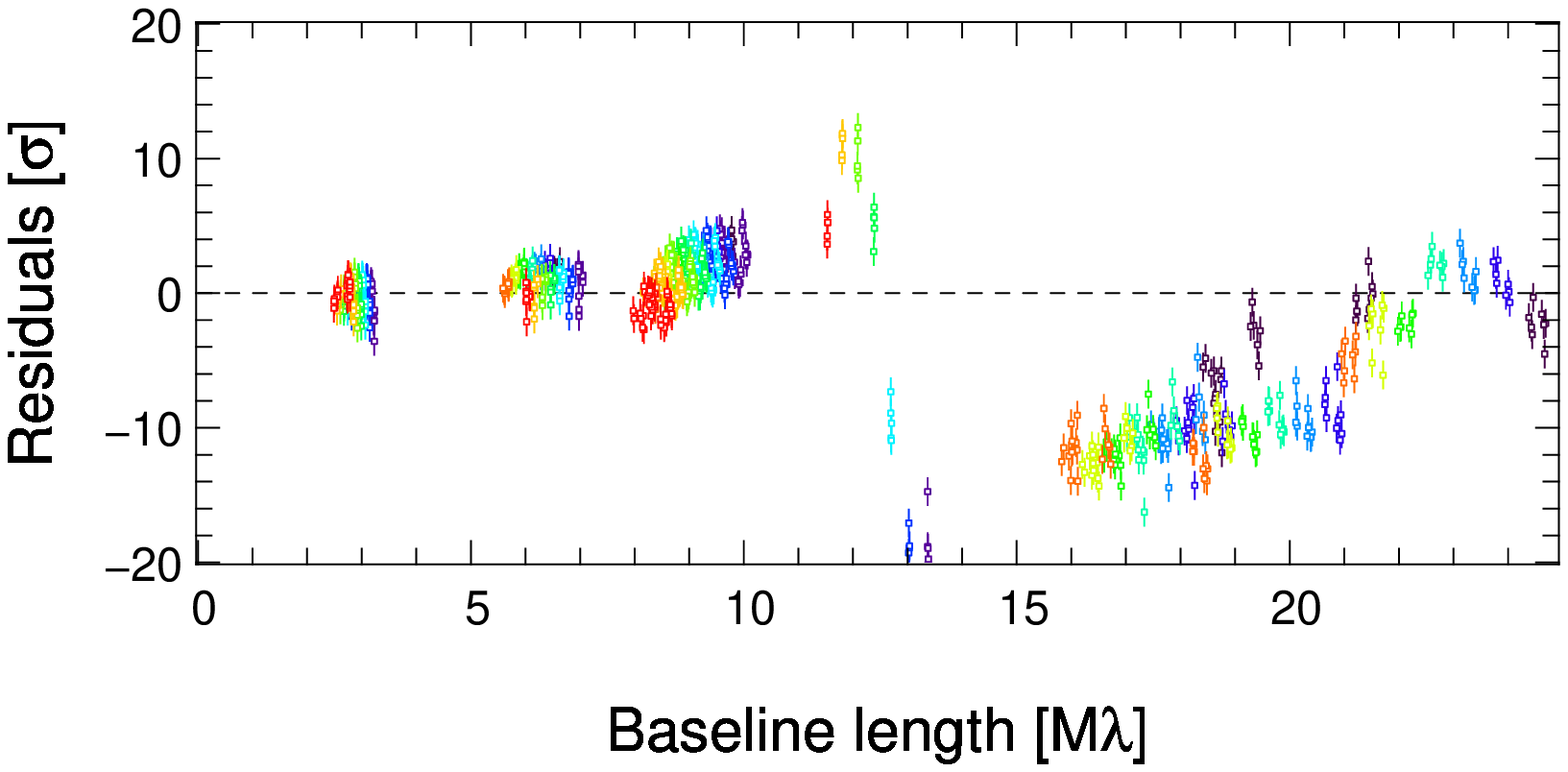}
      \caption{Overview of the dataset: Visibility square measurements
        as a function of the baseline length.  The wavelength is
          color coded with the same palette used in
          Fig.~\ref{fig:UV_planes}. The solid curve correspond to the
        visibility curve of a uniform stellar disk of angular diameter
        20.30 mas (not accounting for bandwidth smearing). The bottom
        panel presents the residual of that fit, showing the clear
        inconsistency of the second lobe.}
         \label{fig:Dataset}
   \end{figure}

   \begin{figure}
   \centering
   \includegraphics[scale=.47]{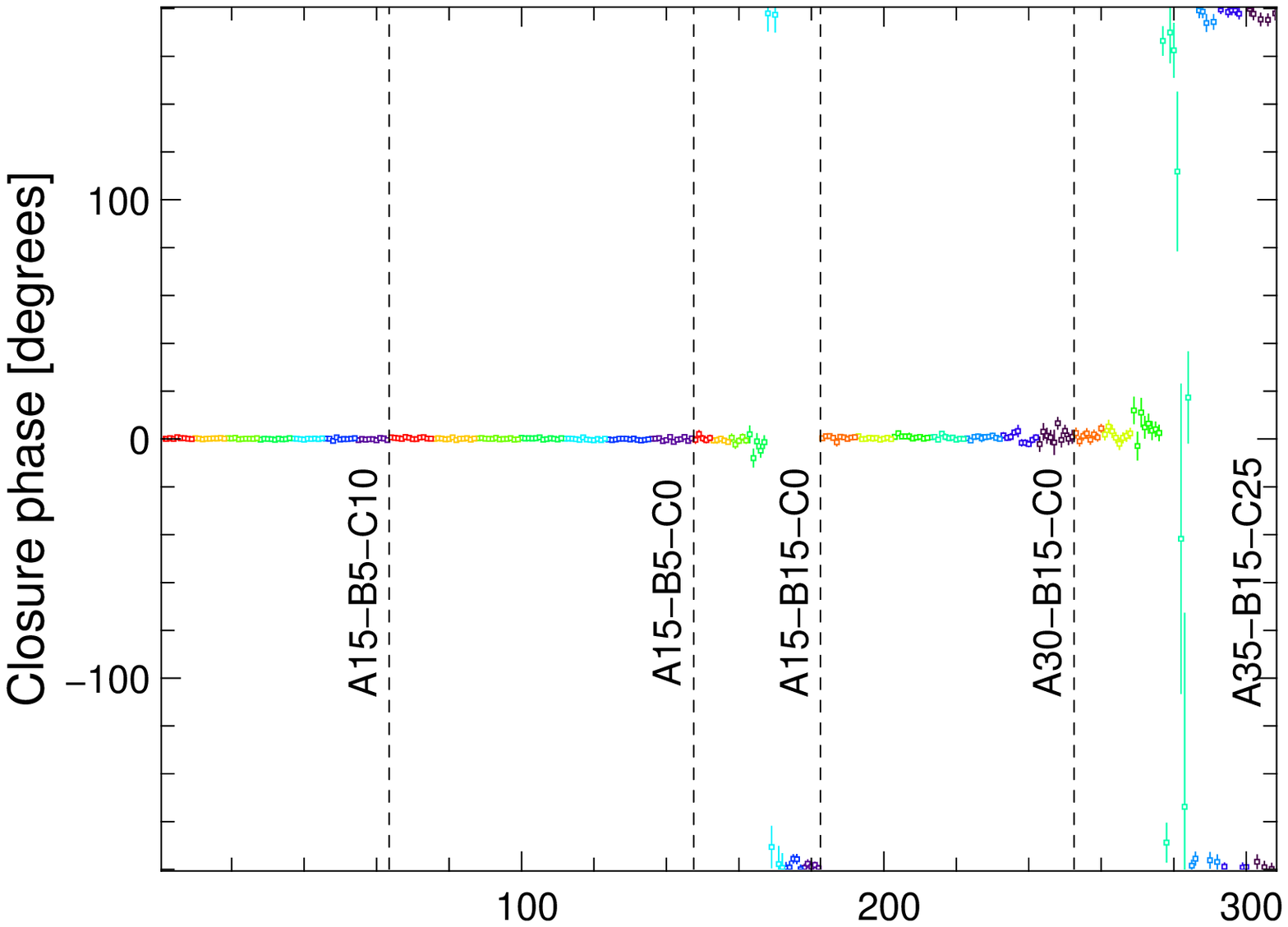}
   \includegraphics[scale=.47]{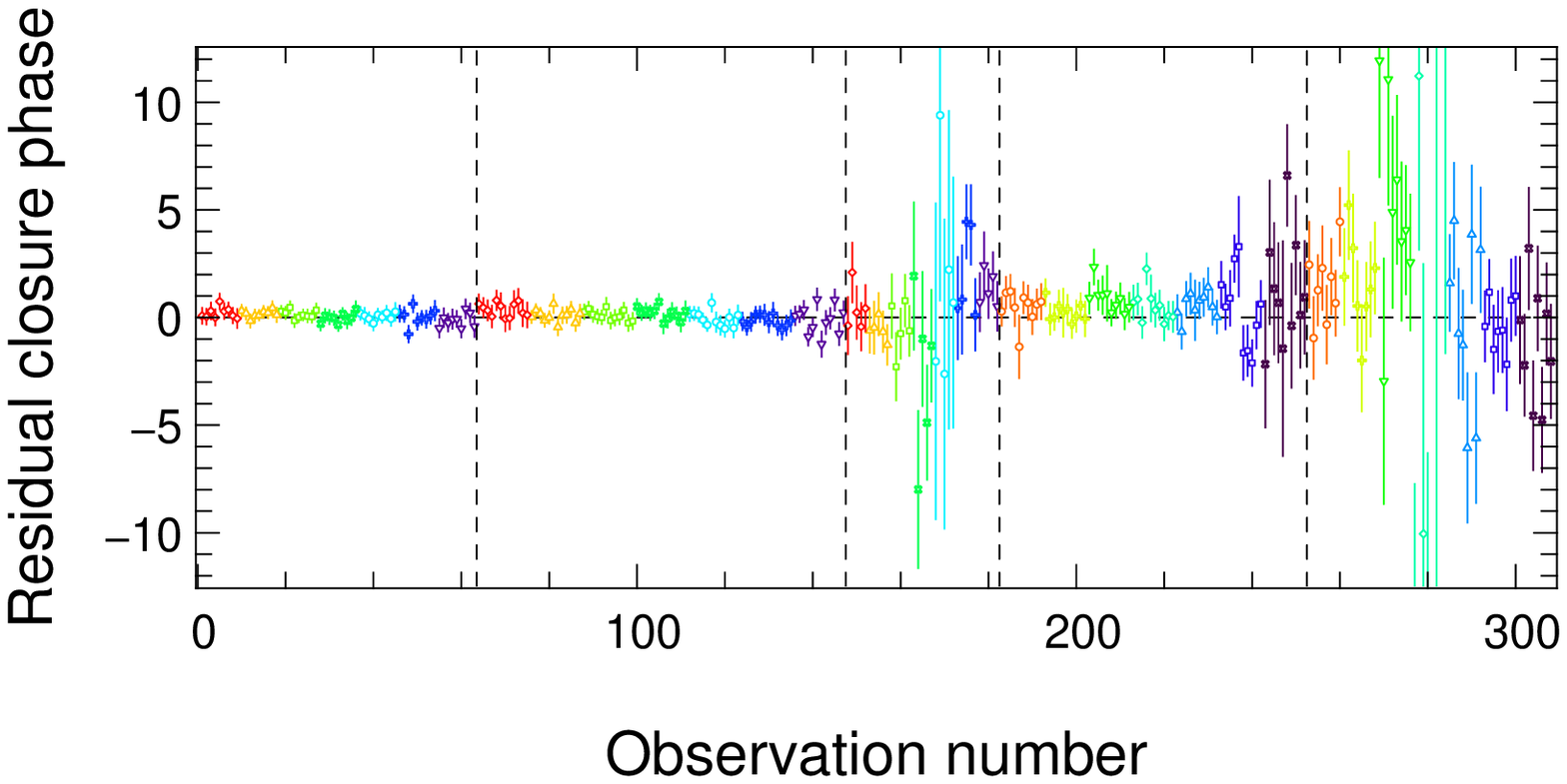}
      \caption{Overview of the dataset: closure phase measurement as a
        function of observation datafile number. The bottom panel
        shows the residual of fitting a simple limb darkening disk to
        the data (power law with parameters as stated in
        Table~\ref{tb:param}). Color legend is similar to the one used
        in Fig.~\ref{fig:UV_planes} and~\ref{fig:Dataset}. Because
        such a model corresponds to a symmetrical brightness
        distribution, the closure phases are either 0 or 180
        degrees. The reduced $\chi^2$ of the closure phase alone is
        1.06.}
         \label{fig:Dataset2}
   \end{figure}

Spectral information was obtained by the means of a prism placed
between the integrated optics and the PICNIC camera
\citep{2003SPIE.4838.1225R}. The temporally-modulated fringes are
therefore spatially dispersed on the detector. To ensure well-defined
spectral edges, we also inserted a broad band H filter in the optical
path. Its bandpass is spatially equivalent to seven pixels on the
camera.

Wavelength calibration of the spectral channels is however a difficult
and critical step. This is especially important since the prism was
removed and re-inserted (with a slightly different position) between the
night of the 13th and the 14th. Fortunately, the spectral wavelength
is coded in the data (see Fig.~\ref{fig:fringes}). The fringe
frequency (in pixels$^{-1}$) is directly proportional to the
wavenumber. The factor of proportionality is constant since the
modulation of the optical path is done by moving the piezo mirrors a
certain distance (step-like) between each pixels reads, even though
the steps are smoothed out by mirror/mount inertia. The relative
wavelength between each channel and each night was established this
way with a precision better than $0.1\%$. This level of precision was
achievable thanks to small differential piston variations due to good
seeing conditions and fast reading mode.  
The relative wavelength between each baseline was also studied. To do
so, we compared the fringe frequency observed at a given spectral channel
between the three baselines. The frequency of the third baseline is
equal to the sum of the frequency of the two first, within $0.2\%$
error bars. This means that the different baselines are at equal
wavelength at a $0.2\%$ level.

However, absolute calibration requires to know the exact angle of the
incoming beam on the piezo mirror. March 2007 narrow band observations
were used, and allowed to establish the speed of the optical path
modulation at $0.188\pm0.002\,\mu$m/sample for the first delay line,
and $0.195\pm0.002\,\mu$m/sample for the second delay line. 
  Optical path modulation was measured on the third baseline at
  $0.383\pm0.004\,\mu$m/sample.  A time sample correspond to the
integration time between two reads. The $\approx1\%$ error bars are
mainly due to uncertain changes in the angle of reflection which
occurred between March and May 2006.

Figure~\ref{fig:wav} summarizes the wavelength calibration results by
plotting the average integrated flux on each spectral channel as a
function of wavelength.  The fairly large error bars in wavelength are
mainly due to the uncertainty on the angle of reflection, and
correspond to a possible global shift in absolute calibration. In contrast, relative wavelength is precisely established, and shows that
a significant displacement of the prism occurred between the night of
the 13th (squares), and the night of the 14th (cross).

\subsection{IOTA field of view}
\label{sc:FOV}

The high resolution of IOTA has a counterpart: the field of view is
limited.  The first limitation is due to the field of view of each
individual telescopes, delimited by the cone of acceptance of the
fibers on the sky. Such value is difficult to estimate, since it
depends on the interferometer as well as the atmospheric seeing. A
first order estimation is to neglect the atmospheric turbulence and to
consider the fiber core to be filling the diffraction pattern of the
telescopes. In this assumption, the field of view of the telescopes
reads:
\begin{equation}
FOV_{\rm telescopes}=\frac{\lambda}{D}\,,
\end{equation}
where D is the diameter of an individual telescope.

A second limitation is the field of view of the interferometer. It is
delimited by the maximum distance between two objects which fringes
overlap on the detector.  To be rigorous, one should take into
account parameters like the mode of recombinaison, the stroke of the
piezo (in the case of IONIC), and even the spectral energy
distribution of the target. However, to establish a simple relation,
we will only take into account the spectral bandwidth of a spectral
channel ($\Delta\lambda$), as well as the distance between two telescopes ($B$):
\begin{equation}
FOV_{\rm interferometer}=\frac{\lambda^2}{\Delta\lambda\,B}\,.
\end{equation}
 Note that the interferometric field of view is baseline dependent. It
  will be larger for shorter baselines, and smaller for longer
  baselines. Moreover, this field limitation is valid only in the
  direction along the baseline. Perpendicular to the baseline, the
  bandpass does not cause any field limitation. All in all, it is
  difficult to establish the field of view of an interferometer as a
  whole.  A conservative way to do so is to consider the maximum
  baseline length for a given direction.

In the North-East/South-West direction, using the spectral dispersion
mode of IOTA ($D=45$\,cm, $\Delta\lambda=40$\,nm and $B=35$\,m), the
field of view is not limited by the telescope ($FOV_{\rm
  telescopes}=750$\,mas), but by the bandwidth.  The field of view is
350$\,$mas at 1.55\,$\mu$m, and 480$\,$mas at 1.80\,$\mu$m. In the
North-West/South-East direction, the shorter baselines ($B=15$\,m)
allow a larger interferometric field of view, hence a 750\,mas field
limitation due to the telescopes size.
 
We will consider in the following of this paper a 400x750\,mas field
of view for IOTA\footnote{IOTA's field of view decrease when using large band
  filters}.

\subsection{The dataset}

The dataset consist of 924 visibility measurements and 308 closure
phases. The $V^2$ are plotted as a function of the baseline length in
the upper panel of Fig.~\ref{fig:Dataset}. The CP are plotted in the
upper panel of Fig.~\ref{fig:Dataset2}. The frequency plane coverage
was previously presented in Fig.~\ref{fig:UV_planes}. The solid curve
corresponds to the best fit of a uniform disk. The residuals are
plotted on the lower panels. It is interesting to note that fringes
have been observed with a contrast below 1\%.   Such a low
  contrast exists thanks to the dispersive mode, which allows a deep
  first null. If the full H band was observed, the effect of bandwidth
  smearing would have limited the depth of the null to several
  percents \citep{2005ApJ...626.1138P}. Probing into the null was
  possible thanks to bootstrapping, two baselines of sufficient
  contrast being enough to track the fringes on all the baselines.

A few things are striking: first, the object is
relatively achromatic. This can be seen on the residuals of the
$V^2$. Secondly, the second lobe of the data is not well fitted by a
uniform disk. This is due to the presence of limb darkening. Thirdly,
the closure phases are close to zero or $\pi$. It means the object is
likely to be point symmetric.

\section{Comparison with atmosphere models / prescriptions}
\label{sc:Com_model}

\subsection{Fitting limb-darkening prescriptions} \label{sc:ima_param}

\begin{table}
\caption{Diameter and limb darkening measurements}
\label{tb:param}
\centering
\begin{tabular}{llc}
\hline \hline 
Law & \multicolumn{1}{c}{parameters} & Reduced $\chi^2$ \\
\hline 
Uniform & $\theta_{\rm UD} = 20.304 \pm 0.011$\,mas  & 31\\
\hline 
Power & $\theta_{\rm LD} = 20.900 \pm 0.007$\,mas  & 1.962\\
 &  $\alpha = 0.258 \pm 0.003$ \\
\hline 
Quadratic &  $\theta_{\rm LD} = 20.922 \pm 0.036$\,mas  & 1.959\\
 &  $a = 0.186 \pm 0.021$ \\
 &  $b = 0.298 \pm 0.053$ \\
\hline 
Quadratic &  $\theta_{\rm LD} = 20.931 \pm 0.004$\,mas  & 2.956\\
Claret (2000)
 &  $a = 0.0291$ \\
 &  $b = 0.5107$ \\
\hline 
Non-linear &  $\theta_{\rm LD} = 20.863 \pm 0.004$\,mas  & 2.013\\
Claret (2000)
 &  $a_1 = 0.8175$ \\
 &  $a_2 = 0.0827$ \\
 &  $a_3 = -0.4116$ \\
 &  $a_4 = 0.1864$ \\
\hline 
{\sc marcs} model &  $\theta_{\mathrm{Ross}}=21.05\pm0.01$\,mas  & 2.080\\
\hline 
\end{tabular}
\begin{list}{}{}
\item[Note --] { Errors bars are pure calculations based on the
  second derivate of the $\chi^2$. They are not valid when assuming an
  unrealistic model of the brightness distribution (for example a
  uniform disk). Moreover, diameter errors do not include the 1\%
  uncertainty due to an eventual wavelength miscalibration (see
  section~\ref{sc:wave}).}
\end{list}
\end{table}

Since limb darkening is apparent, a logical first step is to fit a
model for the brightness distribution of the photosphere. Numerous
types of limb-darkening (LD) prescriptions exist in the literature. We
used two of them, which we supposed achromatic. A power law
\citep{1997A&A...327..199H}:
\begin{equation}
I(\mu)/I(1)=\mu^\alpha\,,
\label{eq:polaw}
\end{equation}
and a quadratic law \citep{1977A&A....61..809M}:
\begin{equation}
 I(\mu)/I(1)=1-a(1-\mu)-b(1-\mu)^2 \,,
\label{eq:quadlaw}
\end{equation}
where $\mu=\sqrt{1-(2r/\theta_{\rm LD})^2)}$, $r$ being the angular distance from
the star center, and $\theta_{\rm LD}$ the angular diameter of the photosphere. In
terms of complex visibilities, the power law limb darkening prescription
 yields:
\begin{equation}
V(v_r) = \sum_{k \geq 0} 
\frac{\Gamma(\alpha/2+2)}{\Gamma(\alpha/2+k+2)
  \Gamma(k+1)} \left( \frac{- (\pi v_r \theta_{\rm LD})^2}{4} \right)^k\,,
\label{eq:polaw_V}
\end{equation}
where $v_r$ is the radial spatial frequency and $\Gamma$ the Euler function (
$\Gamma(k+1)=k!$). On the other hand, the quadratic law yields:
\begin{equation}
V(v_r) =\frac{\displaystyle(1-a-b) \frac{J_1(\zeta)}{\zeta}+ \frac{a+2b}{\sqrt{2/\pi}}
\frac{J_{3/2}(\zeta)}{\zeta^{3/2}}-2b\frac{J_2(\zeta)}{\zeta^2}}{
\displaystyle {1/2}-{a/6}-{b/12}}
\end{equation}
where $\zeta=\pi v_r \theta_{\rm LD}$, $J_1$ and $J_2$ are the first and second-order Bessel function respectively, and:
\begin{equation}
J_{3/2}(\zeta)=\sqrt{\frac{2}{\pi \zeta}} \left(\frac{\sin(\zeta)}{\zeta}-\cos(\zeta)\right) \,.
\end{equation}

Results for the fits are presented in Table~\ref{tb:param}. Using
Eq.~(\ref{eq:polaw}), we obtained a $\chi^2$ of 2413, for 1230 degrees
of freedom. The reduced $\chi^2$ ($\chi^2$ over the number of degrees
of freedom) does not improve significantly when using a two-parameter
prescription for limb darkening, prompting us to consider the power law
model as a sufficient approximation. 

There are two main explanations for the reduced $\chi^2$ different
from one: (i) an underestimation of the error bars, and (ii) an
inexact prescription of the brightness distribution by
Eqs.~(\ref{eq:polaw}) and~(\ref{eq:quadlaw}). Making the distinction
between these two is difficult. On the one hand, photometric
variations of the star are observed at the order of one percent
\citep{2003ApJ...591L.151R}, an indication that the brightness
distribution should not be as smooth and symmetric as our
prescriptions are. On the other hand, no deviation from point symmetry
is observed in the closure phases (section~\ref{sc:comp}) whose
reduced $\chi^2$, taken independently, is 1.015.  The departure from
simple LD models could therefore only be explained by a missing
point-symmetric component. The residuals are discussed more throughly
in the last paragraph of Section~\ref{sc:model}.

Whatever the cause, we decided to be as conservative as possible
  by scaling the errors to a $\chi^2$ of one.  The error bars stated
in Table~\ref{tb:param} are obtained by this mean.  To decrease the
$\chi^2$, we explored -- and discarded -- two other alternatives. The
first one was to increase the error due to calibration (higher than
the 2\% justified in section~\ref{sc:data_red}). However, this
dramatically increased the errors bar on the lowest frequencies, which
is not desired since they are already well fitted by our
prescriptions. The second approach was to add an additive error due to
a potentially imperfect subtraction of the power spectrum bias. Such
an error was included at a 2\% level for faint objects by
\citet{2006ApJ...647..444M}. However, this dramatically increased the
errors bars on the lowest visibilities, which brings an unnecessary
bias on data whose first zero is already well fitted. In conclusion, a
global scaling of the error bars was seen as the best alternative.

\subsection{Fitting the \textsc{Marcs} model}
\label{sc:model}

\begin{figure}
\centering
  \resizebox{\hsize}{!}{\includegraphics{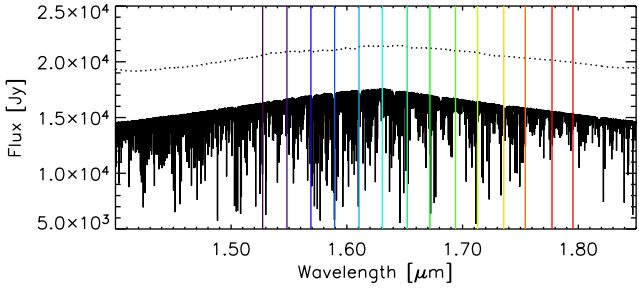}}
  \caption{The synthetic H-band spectrum of the {\sc marcs} model
    (solid line: in opacity sampling resolution, dotted line:
    convolved to the instrumental spectral resolution -- shifted up by
    five thousands Jensky) and the central wavelengths of the spectral
    channels of IONIC.  The peak in the spectrum corresponds to the
    H$^-$ opacity minimum.}
  \label{fig:MARCSspectrum}
\end{figure}  
\begin{figure}
\centering
  \resizebox{\hsize}{!}{\includegraphics{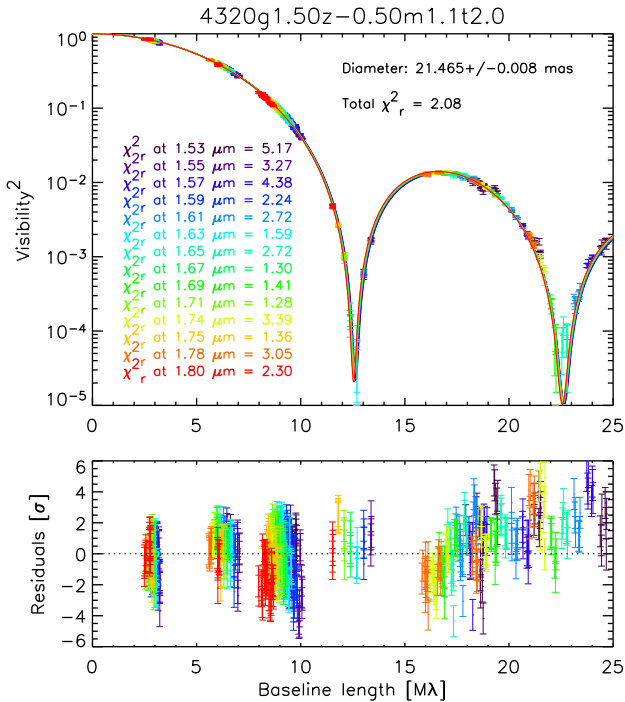}}
  \caption{ Wavelength dependent visibility curves derived from the
    \textsc{Marcs} atmospheric model. The color coding is similar to
    the one used in Fig.~\ref{fig:MARCSspectrum}. IONIC data are
    superimposed on the curves, and residuals are plotted in the lower
    panel. The closure phase residuals are identical to
    those of Fig.~\ref{fig:Dataset2}.  }
  \label{fig:MARCSfit}
\end{figure}

The \textsc{Marcs} atmospheric model was presented in
\citet{2005A&A...435..289V}. The models were originally constructed
and fine-tuned for the calibration of the ISO-SWS (Infrared Space
Observatory Short Wavelength Spectrometer) and checked against FTS
spectra \citep{Decin_these}. For the present study, we searched the
full Arcturus FTS spectral atlas \citep{1995PASP..107.1042H} in the H
band for peculiar spectral features. Lines are sparse and well
spaced. They belong mainly to CN, OH and some atomic transitions. The
IONIC data are therefore ideal to study the H$^-$ continuum, which has
its minimum (the transition between bound-free and free-free regimes)
within the bandpass sampled by our data. The only free parameter to
match model to observations is the angular diameter corresponding to
the outermost point in our model intensity profiles
($\tau_{\mathrm{Ross}} = 10^{-7}$).  Several models with stellar
parameters around those determined by \citet{2003A&A...400..709D} were
used, but they bring no significant improvement in $\chi^2$ compared
to the spectroscopically preferred model (T$_{\rm{eff}} = 4320\,$K,
$\log{g} = 1.5$, [Fe/H] = -0.5 and v$_{\rm{turb}} = 2$\,km\,s$^{-1}$).

The synthetic H-band spectrum calculated from our model and the
comparison of our dataset with this model are shown in
Fig.\,\ref{fig:MARCSspectrum} and Fig.\,\ref{fig:MARCSfit}. We find
the best agreement for a diameter of $21.465\pm0.008$\,mas, which
corresponds to a $\tau_{\mathrm{Ross}} = 1$ diameter of
21.05\,mas. This diameter is slightly larger than the one found in
Sect.\,\ref{sc:ima_param}: the star appears a little smaller at
wavelengths of minimal photospheric opacity than at the
Rosseland-averaged opacity. 

With a $\chi^2$ of 2, this fit is almost as good as it was when
fitting a free-parameter limb darkening prescription. This is an
overall confirmation of the validity of the \textsc{Marcs} modeling of
the limb darkening. Analysis of the visibility residuals plotted in
Fig.~\ref{fig:MARCSfit} indicates some possible shortcomings of the
model (supposing error bars are not underestimated, see
Section~\ref{sc:ima_param}). Indeed, the high $\chi^2$ can be
accounted for by two biases: a chromatic bias at low frequency
($\approx9\,$M$\lambda$), another achromatic at high frequencies
(around the second nul). Accounting for these biases could be done by
(i) introducing a circumstellar emission of H$_2$O at a level of half
a percent \citep[water detection was reported by][although normal
  hydrostatic model does not predict any in the
  photosphere]{2002ApJ...580..447R}, and (ii) slightly modifying the limb
darkening distribution. However, such possibilities are at the limit
of what we think is reasonable to derive from our data, and no further
modeling was done to avoid over-interpretations.

\subsection{On the angular diameter of Arcturus}

Numerous angular diameter measurements can be found in the
literature. Previous interferometric observations either use
uniform-disk fitting and apply limb-darkened corrections, or fit disks
whose limb-darkening is fixed by atmospheric models. At a wavelength
of $2.2\mu$m, \citet{1986A&A...166..204D} observed Arcturus with the
I2T interferometer and published a diameter of $\theta_{\rm UD}
=20.36\pm0.20$\,mas as well as a limb-darkened value $\theta_{\rm LD}
=20.95\pm0.20$\,mas. Previous measurements using the IOTA
interferometer exist too, and yielded in the K band $\theta_{\rm LD}=
19.5\pm1.0$\,mas \citep[][using bulk optics]{1996AJ....111.1705D},
$\theta_{\rm LD}=20.91\pm0.08$\,mas \citep[][using
  FLUOR]{1998A&A...331..619P} and $\theta_{\rm
  Ross}=21.18\pm0.21$\,mas \citep[][also using
  FLUOR]{2005A&A...435..289V}. In the visible,
\citet{2003AJ....126.2502M} observed Arcturus using the MarkIII
interferometer (450-800\,nm), and after correction for a substantial
limb darkening effect, published $\theta_{\rm LD}
=21.373\pm0.247$\,mas.

From our dataset, and taking into account wavelength calibration
uncertainties, we derived $\theta_{\rm LD} =20.91\pm0.21$\,mas and
$\theta_{\rm Ross} =21.05\pm0.21$\,mas. These results are in agreement
with I2T, MarkIII and IOTA observations \citep[$1.5\sigma$ in][]{1996AJ....111.1705D}. It does not yield
an increase in terms of precision, but our measurements are indeed
interesting since, unlike the others, they did not require a
pre-defined value to account for limb darkening. Using the Rosseland
diameter and \citet{1999AJ....117.2998G} estimation of the integrated
flux ($F=(4.98\pm0.02)\times10^{-5}$\,erg\,cm$^{-1}$\,s$^{-1}$), we
can update their calculation of Arcturus' effective temperature to
$T_{\rm eff}=4295\pm26$\,K.

\subsection{On the limb darkening}

\begin{figure}
\centering
  \resizebox{\hsize}{!}{\includegraphics{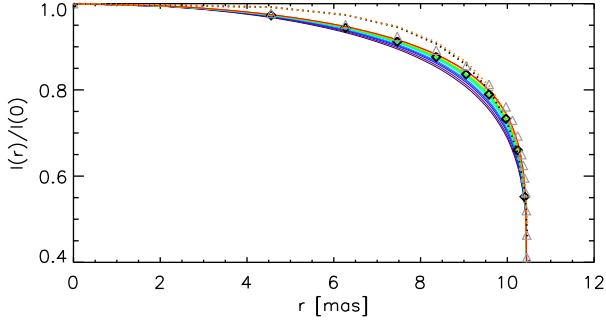}}
  \caption{Intensity profiles of our {\sc marcs} model as colored
    lines. For comparison, the best parametric fit of a power law is
    represented with diamonds, the Kurucz model with triangle, and the
    quadratic LD curve \citet{2000A&A...363.1081C} as a dotted
    line. This last fit differs significantly form the others,
    revealing a problem in the method used for limb darkening fitting
    used by Claret.}
  \label{fig:IP}
\end{figure}  

%
\begin{figure}
\centering
  \includegraphics[width=9cm]{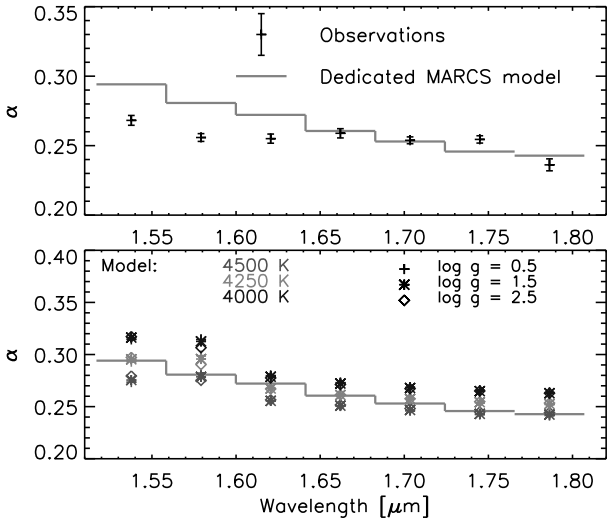}
  \caption{Limb darkening as a function of the wavelength.  {\it
        Upper panel}: the observed LD coefficients $\alpha$ (with error bars)
      are compared with the LD coefficients derived by fitting a
      $I(\mu) = \mu^{\alpha}$ profile to the {\sc Marcs} intensity
      profiles of our dedicated Arcturus model (T$_{\rm{eff}} =
        4320\,$K, $\log{g} = 1.5$, 
      [Fe/H] = -0.5 and v$_{\rm{turb}} = 2$\,km\,s$^{-1}$).  {\it
        Lower panel}: influence of temperature and gravity on the limb
      darkening. 
      }
  \label{fig:alphas}
\end{figure}  
An important prospect of this work was to compare our limb darkening
measurements with existing atmospheric models. A first test was to
derive parameters of the limb darkening, and compare them with
published values in the literature. We were greatly surprised to see a
strong inadequacy between our measurements and the quadratic
parameters given by \citet{2000A&A...363.1081C} (see
Table~\ref{tb:param} and Fig.~\ref{fig:IP}).  However, they also published
the values for a more complex 4-parameter non-linear law:
\begin{equation}
 \frac{I(\mu)}{I(1)}=1-a_1(1-\sqrt{\mu})-a_2(1-\mu)-a_3(1-\mu^{3/2})-a_4(1-\mu^2) \,.
\label{eq:non_linear}
\end{equation}
\citet{2000A&A...363.1081C} claims that this four-parameter non-linear law
should give a more reliable estimation of the limb darkening. Using
his published parameters (assuming T$_{\rm{eff}} = 4250\,$K, $\log{g}
= 1.5$, [Fe/H] = -0.5 and v$_{\rm{turb}} = 2$\,km\,s$^{-1}$) we were
able to indeed confirm a correct fit.

   \begin{figure*}
   \centering
   \includegraphics[width=7cm]{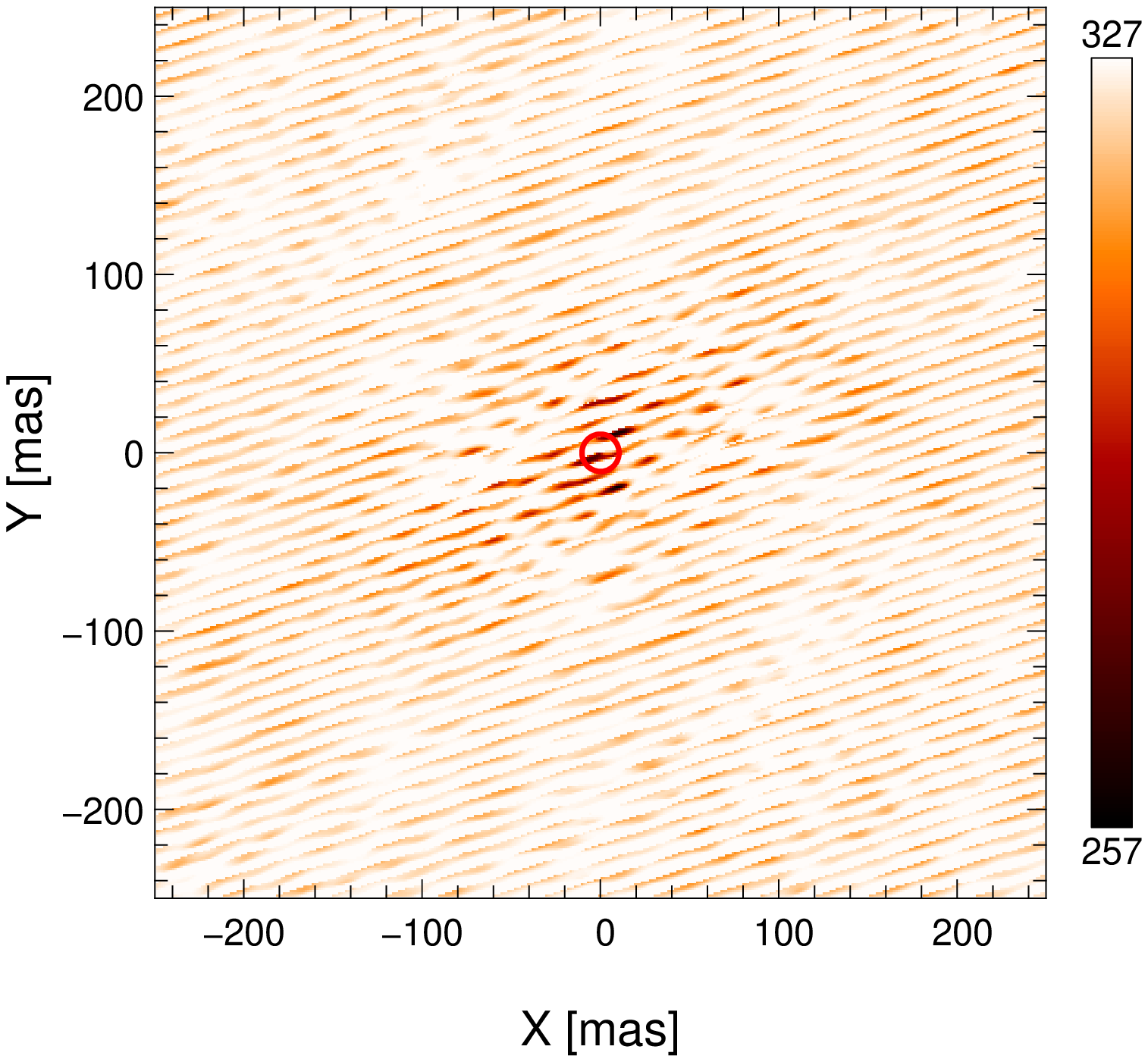}\hspace{1cm}
   \includegraphics[width=7cm]{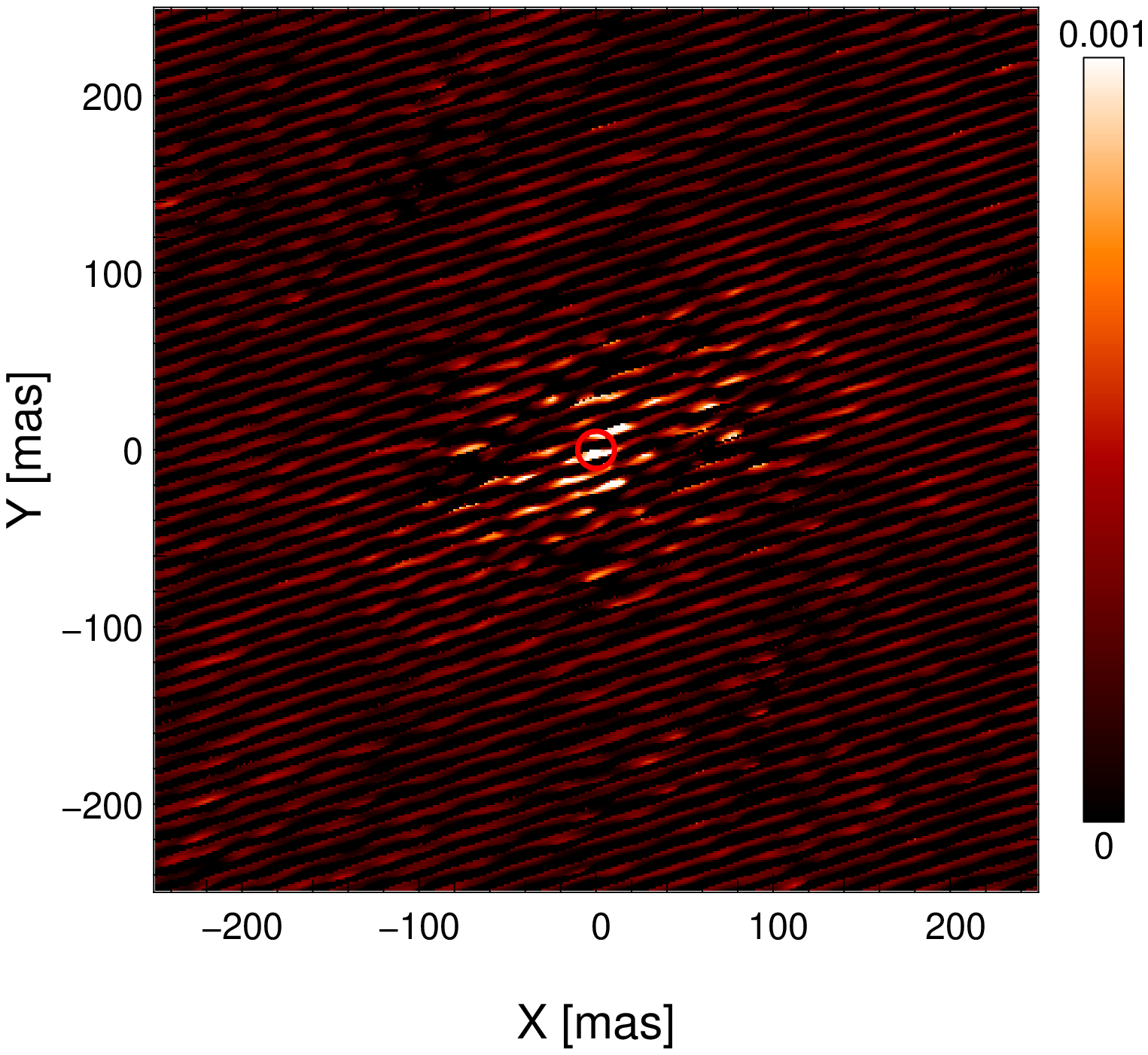}
      \caption{Non-detection of a companion to Arcturus. Left panel:
         closure phase $\chi^2$ map. Right panel: $1\sigma$
        upper limits of the brightness ratio of a companion. The
        circles represent the limit of the photosphere. The field of
        view of IOTA is 400x750\,mas (see discussion in
        Section~\ref{sc:FOV}).}
         \label{fig:Comp}
   \end{figure*}

However, it is not the quadratic law which is intrinsically less able
to match the limb darkening: when leaving the parameters free to
adjust, the $\chi^2$ of a quadratic law is able to match the $\chi^2$
of the non-linear law. Therefore, the problem with the quadratic
values published by \citet{2000A&A...363.1081C} should lie in the
method to derive the parameters. To confirm this, we used the
\textsc{Atlas} model \citep{1979ApJS...40....1K} -- the one used by
Claret -- and we were able to obtain a good fit for the limb darkening
(see Fig.~\ref{fig:IP}). An explanation could be found in a paper
written by \citet{2007ApJ...656..483H}, in which the author states that
conventional stellar limb fitting methods (like the one used by
Claret) are biased.

But the most striking results from Table~\ref{tb:param} is the
consistency in fitting quality achieved when using different limb
darkening laws (except when using Claret's quadratic value). We noted
that both \textsc{ Marcs} and \textsc{Atlas} models give similar fits,
showing an equivalent capacity to correctly model the atmosphere of
Arcturus. The reduced $\chi^2$ values are not exactly 1, but are close
to the ones obtained when fitting LD laws with freely variable
parameters. This is a good validation of both atmospheric modeling
softwares.   Secondly, we do not note any difference in the
  fitting quality between a power and a quadratic limb darkening
  law. Furthermore, the likelihood does not increase when using a
  4-coefficient non-linear law (reduced $\chi^2$ of 1.97 -- we do not
  present the results in Table~\ref{tb:param} since none of the
  coefficients are properly constrained by our dataset).  This is
  because we do not have the necessary angular resolution to
  distinguish the several limb darkening laws used here. To our
  dataset, all of them are equally good. Therefore, for angular
  resolution no greater than ours, we recommend using the power law
  instead of the two other limb darkening laws tested in this work,
  since it would use less free parameters while still being able to
  correctly model the LD.

Finally, we investigated the spectral dependence of the limb
darkening. To do so, we fitted a limb-darkening power law with a
wavelength-dependent $\alpha$ value to the observations. The fit was
done using an achromatic diameter as, in principle, there is no such
thing as a different diameter at different wavelengths: a different
intensity profile, or in the case of a well-behaved star just a
different LD, mimics a different diameter at different wavelengths. In
the absence of extended molecular layers and other similar large
deviations from a normal photospheric IP, this effect is mostly
accounted for by the LD parameter.  Similarly, we derived theoretical
$\alpha$ values from our preferred \textsc{marcs} model.  The result
is summarized in Fig.~\ref{fig:alphas}. The general agreement is quite
good.  The almost linear slope of the limb-darkening is in fact a
complex combination of opacity due to the H${^-}$ continuum and
molecular absorptions features.  A minor discrepancy is an
overestimation of the LD at the blue end of the bandpass. We searched
a grid of models for possible improvement, but no significantly better
fit could be attained with reasonable stellar parameters.

\section{On the point symmetry of Arcturus}\label{sc:comp}

\subsection{Fitting the closure phases}

Closure phases (CP) are extremely sensitive to deviations from point
symmetric brightness distributions. For example, a binary of contrast
ratio 1:100 could induce closure phases of several tens of degrees at
low visibilities. The mean of our first 147 CP measurements (first two
days of observation) is 0.067 degree, with an average root mean square
of 0.34 degree. Such high quality data is therefore excellent for
probing a companion. When fitting a power law limb-darkened disk to
the data (both $V^2$ and CP; see section~\ref{sc:ima_param}), the
$\chi^2$ on the CP was 327 over 308 closure phases --- corresponding
to a reduced $\chi^2$ of 1.07. Fig.~\ref{fig:Dataset2} shows the CP as
well as the residual of the fit. The fit is, in our opinion,
satisfactory.

An upper limit for the brightness ratio of a possible companion can be
obtained. We modified the visibility function presented in
Eq.~(\ref{eq:polaw_V}) to account for the presence of a point-like
off-centered source:
\begin{eqnarray}
V(u,v) &=& (1-K) \sum_{k \geq 0} 
\frac{\Gamma(\alpha/2+2)}{\Gamma(\alpha/2+k+2)
  \,k!} \left( \frac{- \pi^2 \theta_{\rm LD}^2(u^2+v^2)}{4} \right)^k \nonumber \\
&&+ K \exp\left(2i\pi(Xu-Yv)\right)\,.
\label{eq:V2_comp}
\end{eqnarray}
$K$ is the brightness ratio of the companion, $X$ and $Y$ its position
and $u$ and $v$ the spatial frequencies (arcsec$^{-1}$). The star
parameters ($\alpha$ and $\theta_{\rm LD}$) are fixed to the value
presented in Table~\ref{tb:param}. For each position of the companion
Eq.~(\ref{eq:V2_comp}) is computed, CP are derived and $K$ is adjusted
to minimize the $\chi^2$ on the closure phases. The minimum $\chi^2$
are plotted as a function of $X$ and $Y$ on the left panel of
Fig.~\ref{fig:Comp}. The general minimum $\chi^2$ for a companion
situated within the field of view of IOTA but further away than 1\,AU
of the star (400\,mas$>\sqrt{X^2+Y^2}>89$\,mas) is 299, with a
brightness ratio $K=(4\pm4)\times10^{-4}$. This is not significant
enough to be considered a detection. The values $(K+\sigma(K))$ can
nevertheless be used to derive upper limits for the brightness ratio
of a possible binary system. It is plotted in the right panel of
Fig.~\ref{fig:Comp}. The average dynamic range at 1\,AU of the star is
$1.5\times10^{-4}$.

Closer to the photosphere, the $\chi^2$ can decrease
substantially. The minimum is 257, for $X=10$ and $Y=11$\,mas. It
still does not mean we have detected anything, since this value is
below the number of degrees of freedom. However, the fit can be used to
put upper limits on the brightness of a possible hotspot on the
photosphere. The maximum value for $K+\sigma(K)$ on the photosphere is
$1.7\times10^{-3}$. Note that the signature of an hotspot on the CP
gets smaller when it gets closer to the photocentre. Therefore, we
cannot exclude the presence of a bright hotspot coincidentally
situated in the middle of the photosphere.

\subsection{Presence of a companion?}

Arcturus is often used for high-resolution spatial and spectral
calibration \citep{2000ApJ...534..907T,2003A&A...400..709D}. Such use
makes this star both very well known and very important to know. This
explains why, when {\it Hipparcos} flagged this star as a binary, it
stirred an important debate in the community. The absence of other
observational evidence \citep{1998Obs...118..299G}, uncertainties in
the {\it Hipparcos} detection \citep{1998Obs...118..365S} and finally
non-detection with adaptive optics observations
\citep{1999PASP..111..556T}, convinced the community they could keep
using Arcturus as a calibrator. Our results put an upper limit on the
brightness ratio of a possible companion to $8\times10^{-4}$ in the H
band.

To make our results compatible with a binary system as proposed by
{\it Hipparcos} \citep{1997A&A...323L..49P} or
\citet{2005A&A...435..289V} ($\Delta m \approx 4$, $\rho \approx 230$\,mas,
$M \geq 0.7 M_\odot$), we would have to imagine either (i) a strong
dependence of the wavelength or (ii) an edge-on orbit with the
secondary occulted by the primary. Both possibilities can be ruled out
since (i) a differing spectral type would have aroused spectroscopists
and (ii) an edge-on orbit would have aroused people doing radial
velocity measurements.  However, a lower mass planet of a few Jovian
masses, as proposed by \citet{1989PASP..101..147I,1993ApJ...413..339H}
and \citet{2007PASP..119..237B} is still a possibility. Our
measurement gives an upper limit on its relative magnitude in the H
band ($\Delta m>7.75$).

\subsection{Asymmetric brightness distribution of the stellar surface}

Radial velocities \citep{1999ASPC..185..187M} as well as photometry
\citep{2003ApJ...591L.151R,2007arXiv0706.3346T} indicate variations of
a few days period. Photometric oscillations are especially notable,
with amplitude variation of up to a percent, well above what is
predicted by atmospheric models \citep{2001MNRAS.328..601D}. By
putting a $1.7\times10^{-3}$ 1\,$\sigma$ upper limit on the flux of an
eventual hotspot, our observations show that the temporal brightness
oscillations do not have a spatial counterpart. It means the source of
these variations is most likely not due to convection cells and/or
non-radial oscillations. It is interesting to note that interferometry
could be a good tool to detect non-radial pulsation in variable stars
($\beta$ Cephei, ...).

\section{Imaging Arcturus} \label{sc:image}

\subsection{The \textsc{Wisard} and \textsc{Mira} reconstruction softwares}

   \begin{figure*}
   \centering \includegraphics[scale=.63,angle=-90]{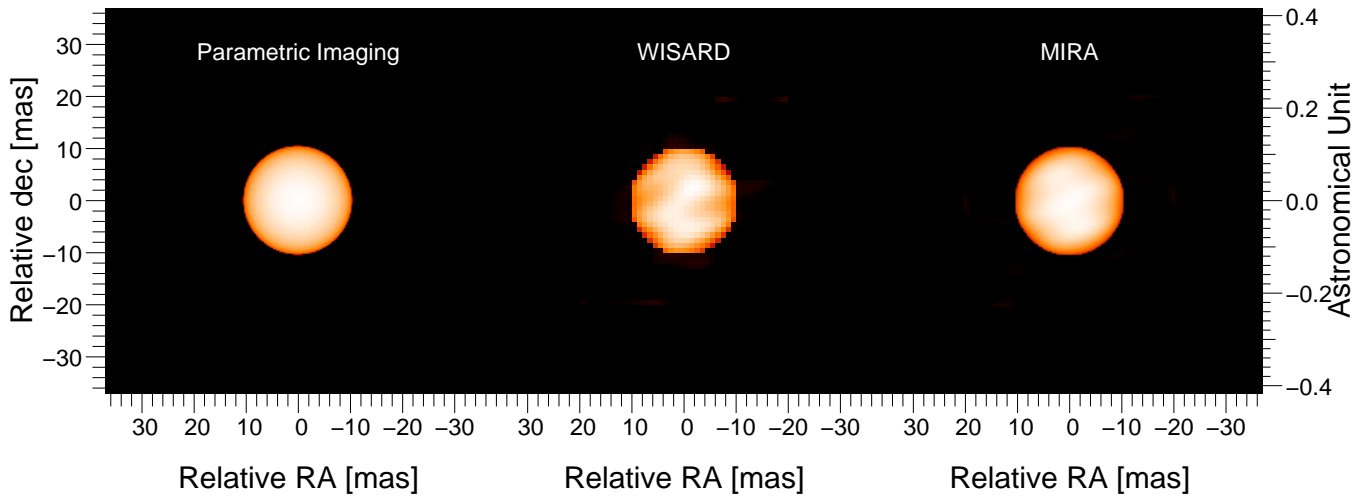}
      \caption{Three different images of Arcturus. On the left hand,
        image produced using a limb-darkening power law with parameter
        from Table~\ref{tb:param}. The two others images are obtained
        using two different softwares for regularized imaging. The
        right hand scale is derived using the {\it Hipparcos} parallax
        of $88.83\pm0.53$\,mas \citep{leeuwen07}.  }
         \label{fig:Images}
   \end{figure*}

Model fitting confines the image to be within the range of a
pre-defined model. This is a perfect tool to derive parameters of
astronomical objects whose morphology is already known. However, it
would not reveal any unexpected phenomenon, hence the need for less
constraining image reconstruction.

The image is sought by minimizing a so-called \emph{cost function}
which is the sum of a regularization term plus data related terms. The
data terms enforce the agreement of the model image with the different
kind of measured data (power spectrum, phase closures, complex
visibilities, etc).  The interpolation of missing data is allowed by
the regularization and by strict constraints such as the positivity
(which plays the role of a floating support constraint) and
normalization.

To validate this imaging process, we used two different reconstruction
algorithms: \textsc{Wisard} and \textsc{Mira}.  \textsc{Wisard}
\citep{Meimon-t-05,Mugnier-l-08a} stands for ``Weak-phase
Interferometric Sample Alternating Reconstruction Device''. Its
approach consists in finding the image and the missing phase data
jointly. This technique is called \textit{self-calibration} in
radio-interferometry \citep{Cornwell-81} and has enabled reliable
images to be reconstructed in situations of partial phase
indetermination. The strength of \textsc{Wisard} is that it combines,
within a Bayesian framework, a recently developed noise model
approximation suited to optical interferometry data
\citep{2005JOSAA..22.2348M}, and an edge-preserving regularization
\citep{2004JOSAA..21.1841M} to deal with the sparsity of the data
typical of optical interferometry.

\textsc{Mira} \citep{Thiebaut_etal-2003-JENAM} stands for
``Multi-aperture Image Reconstruction Algorithm''.  Compared to
\textsc{Wisard}, \textsc{Mira} does not explicitly manage the missing
Fourier phase information: all missing information is handled
implicitly in the data related term. For instance, it is possible to
reconstruct an image given only the Fourier modulus information
\citep[power spectrum data;][]{Thiebaut-2007-GRETSI}. A second
difference -- in these image reconstructions of Arcturus -- lies in
the chosen prior. Instead of an edge preserving regularization, we
used for this reconstruction a quadratic regularization criterion.  To
that end, we computed a prior which is a parametric model image of a
stellar surface (a quadratic limb darkening law). This method, similar
to that used by \citet{Monnier_etal-2007-Altair}, has the
particularity of requiring a rough model of the observed object. It
will inject more information into the reconstruction, which in turn
can give wrong results if the prior model is not right.

We shall note that both \textsc{Wisard} and \textsc{Mira} algorithms
can use various types of priors (entropy, Tikhonov, etc). It is
therefore possible to have both algorithms using the same prior. The
phase management -- explicit or implicit -- will however be different.

\begin{figure}
   \centering \includegraphics[scale=.45]{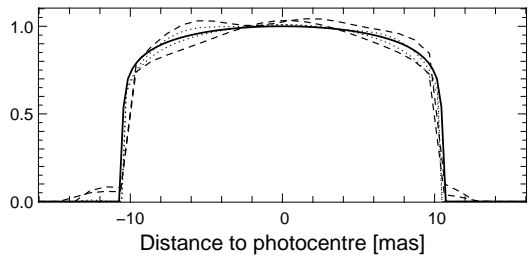}
      \caption{Intensity slices through the $x$ and $y$ axes of the
        three images of Arcturus presented in
        Fig.~\ref{fig:Images}. The solid curve correspond to the
        parametric image, the dashed curves to the \textsc{Wisard}
        reconstruction, and the dotted curves to the \textsc{Mira}
        reconstruction. If we suppose the limb darkening prescription
        to be the correct brightness distribution of the object, we
        can derive the residual of the reconstruction obtained by the
        two imaging softwares: the rms error is around 5\% for the
        \textsc{Wisard} prior (edge-preserving) and 2-3\% for
        \textsc{Mira} (limb-darkening prior).}
         \label{fig:Images_cut}
   \end{figure}

The images reconstructed by \textsc{Wisard} and \textsc{Mira} are
shown in Fig.~\ref{fig:Images}. A third representation of Arcturus is
also presented. It is an image reconstructed from the parameters
derived by fitting a power law limb-darkening prescription on the data
(values presented in Table~\ref{tb:param}). We tentatively call such
type of image reconstruction ``Parametric imaging''. Cuts of the
brightness distributions are presented Fig.~\ref{fig:Images_cut}.  The
similarity of the reconstructions is quite striking considering that
the 2 reconstruction methods (i.e. data-fidelity terms), as well as the
priors, are different.

\subsection{Discussion of image reconstruction}

The left hand image of Fig.~\ref{fig:Images} shows a featureless
limb-darkened star. It is not a surprise since the image is strictly
constrained by the prescription. However, the important result is the
good fit of the prescription on our data. When doing parametric
imaging, the $\chi^2$ is a strong information to judge the reliability
of the image reconstruction. In this case, a reduced $\chi^2$ of 1.9
for 2413 degrees of freedom is a good validation of the derived image.

When dealing with regularized imaging, it is more difficult to judge
the reliability of an image reconstruction. This is because the
quantity which is minimized is a sum of a regularization term and a
data term (generally, the $\chi^2$). The minimum of this {\em cost
  function} is therefore dependent on the regularization term, and no
process is known which could use this minimum to judge on the quality
of the reconstruction. The $\chi^2$ is still of interest, but does
only give a partial view on the reliability of the reconstructed image:
a reduced $\chi^2$ close to one is important, but it is not a
quantity by itself which will ensure the quality of the image
reconstruction.

Ultimately, the quality of image reconstruction will be dependent on
the choice of the regularization term. The closer to the object the
regularization term brings us, the closer to the reality our image
reconstruction will be. According to this philosophy, it is important
to have a good estimation of the prior. We recommend for optical
interferometrist to use an adjustable regularization term.
Simultaneously or sequentially, a solution we propose consists in: (i)
fit a parametric image which best describe the data, and (ii) find the
image which best fit both the data and the parametric image. This
technique was the one used with the \textsc{Mira} reconstruction
software.

\section{Conclusion}
\label{sc:conclusion}

In this paper, we presented an Arcturus dataset of interesting
quality. With the IOTA/IONIC interferometer, we measured fringe
contrasts of less than a percent, with errors bars in average below
that level. Using this data, we fitted several models and
prescriptions. The closure phases were well fitted by point-symmetric
prescriptions. No companion at less than one AU was detected with an
upper limit on its contrast ratio of $8\times 10^{-4}$. The same
modeling of the closure phases allowed the derivation of an upper
limit on the heterogeneity of the photosphere: no hotspot with a
brightness above $1.7\times 10^{-3}$ the total flux of the photosphere
was detected.

We adjusted \textsc{Marcs} atmospheric models to the data. The derived
Rosseland diameter equaled $21.05\pm0.21$ mas, most of the error bar
being induced by non-trivial wavelength calibration. With a reduced
$\chi^2$ of 2, it is interesting to note that atmosphere models of
regular K giants can now be challenged by interferometry at a very
fundamental level, even though spectroscopic agreement is
near-perfect. Interestingly, we noted (i) a slight inconsistency in
the magnitude of the limb-darkening at short wavelength ($\lambda
\approx 1.55\,\mu$m; see Fig.~\ref{fig:alphas}), and (ii) a slight
chromatic effect present in the residual (lower panel of
Fig.~\ref{fig:MARCSfit}). This last results could hint the presence of
a marginal ($\approx 0.5$\%) water vapor emission outside the
photosphere.

Finally, we imaged the photosphere using two different reconstruction
algorithms (\textsc{Wisard} and \textsc{Mira}). Both produced
realistic images, but highlight the difficulty to judge the
reliability of regularized image reconstruction. Comparatively, we
presented an image reconstructed from an ad-hoc prescription of a limb
darkened stellar surface. The low number of free parameters, combined
with a good fit to the data, hinted to us that the most realistic
brightness distribution is in fact the one of a simple limb darkened
disk.

\begin{acknowledgements}
SL acknowledges financial support through a {\it Lavoisier}
fellowship. This work also received the support of PHASE, the high
angular resolution partnership between ONERA, Observatoire de Paris,
CNRS and University Denis Diderot Paris 7.
\end{acknowledgements}

\bibliographystyle{aa}
\bibliography{ABbib,Cygbib}

\end{document}